# Parity-time-symmetric optical lattice with alternating gain and loss atomic configurations

Zhaoyang Zhang[1,2], Liu Yang[3], Jingliang Feng[2], Jiteng Sheng[4], Yiqi Zhang[1], Yanpeng Zhang[1*] and Min Xiao[2,5†]

[1]*Key Laboratory for Physical Electronics and Devices of the Ministry of Education & Shaanxi Key Lab of Information Photonic Technique, Xi'an Jiaotong University, Xi'an, 710049, China*

[2]*Department of Physics, University of Arkansas, Fayetteville, Arkansas, 72701, USA*

[3]*College of Automation, Harbin Engineering University, Heilongjiang 150001, China*

[4]*State Key Laboratory of Precision Spectroscopy, East China Normal University, Shanghai, 200062, China*

[5]*National Laboratory of Solid State Microstructures and School of Physics, Nanjing University, Nanjing 210093, China*

*Corresponding authors:*[*]*ypzhang@mail.xjtu.edu.cn,* [†]*mxiao@uark.edu*

Since the spatially extended periodic parity-time (PT) symmetric potential can possess certain unique properties compared to a single PT cell (with only a pair of coupled gain-loss components), various schemes have been proposed to realize periodic PT-symmetric potentials based on optical lattices. Here, we experimentally construct a spatially periodic PT-symmetric optical potential based on gain-loss arrays induced in a coherently-prepared atomic medium. The gain and loss arrays are generated in alternating four-level N-type and three-level Λ-type configurations in the same atomic medium, respectively, which do not require discrete diffractions as demonstrated in the previous work [*Phys. Rev. Lett.* **117**, 123601(2016)] and can be easier to realize with more relaxed operating conditions. The dynamical behaviors of the system are investigated by measuring the phase difference between two adjacent gain and loss channels. The demonstrated PT-symmetric optical lattice with easy accessibility and better tunability sets a new stage for further exploiting the peculiar physical properties in periodic non-Hermitian systems.

The studies on non-Hermitian Hamiltonians with entirely real eigenvalue spectra have made great progresses in the past decade since the introduction of parity-time (PT) symmetry concept [1, 2] into optics, considering the formal equivalence between the Schrödinger equation in quantum mechanics and the paraxial wave propagation equation in optics[3, 4]. Such equivalence makes complex PT-symmetric potential effectively achievable in an optical setting by spatially arranging the real and imaginary parts of the refractive index in even and odd symmetric manners, respectively. The demonstrations of exact PT symmetry have been mainly made in pairs of coupled optical components (such as microresonators and waveguides) with respective gain and loss [5]. Given that single PT cells (one pair of gain-loss elements) have demonstrated many unconventional optical properties [6-11], interesting exotic features are expected in the extended non-Hermitian optical lattices [12-34]. The proposed behaviors for light travelling through non-Hermitian lattices spread over non-Hermitian optical solitons [14-17], non-Hermitian Bloch oscillation [18, 19], unidirectional invisibility [21, 22], PT-symmetric Talbot effect [27], and nonreciprocal characteristics [30], just to name a few. Studying these novel effects in periodic non-Hermitian optical settings may provide new routes for exploring useful applications in non-Hermitian synthetic materials and further constructing on-chip optical integrated devices.

So far, experimental realizations of periodically arranged PT-symmetric lattice potentials had been limited only to a few optical settings, including passive waveguides with periodically modulated dielectric permittivity [22], a time-domain lattice realized in a network of coupled gain/loss fiber loops [19, 23], and an optical waveguide array with a lossy (without gain) background [24]. The only known exact PT-symmetric optical lattice with spatially modulated periodic gain and loss channels is realized in a four-level N-type atomic configuration [35]. In that experiment, two pairs of laser beams (coupling and pump fields) form the lattice with alternative gain/loss channels in the atomic medium, and a Gaussian signal beam travelling through the induced optical lattice experiences discrete diffractions from this spatially modulated PT-symmetric lattice. Such observation in solid materials is not easy to do due to various restrictions in material properties including the limitation on engineering desired periodic gain in certain materials as well as the connection between the real and imaginary parts of the index as imposed by Kramers-Kronig relations [36]. Different from solid-state systems, atomic media are quite easy and efficient in constructing desired real/imaginary refractive index profiles [37]. By making use of the laser-induced atomic coherence, particularly with the electromagnetically induced transparency (EIT) [38-40] technique, one can easily construct controllable and desired linear (dispersion, gain/loss) and nonlinear properties in coherently-prepared multi-level atomic media [41].

In this paper, we experimentally demonstrate a new scheme to construct a spatially extended PT-symmetric optical lattice based on the periodically alternating gain and loss induced in respective four-level N-type and three-level Λ-type atomic configurations, which appear alternately along the transverse $x$ direction in the same $^{85}$Rb atomic medium. The four-level system providing Raman gain is driven by a signal field, a coupling field and a pump field. With the intensity of the pump field set to zero, the four-level system reduces down to a three-level one, which provides a modifiable loss under the EIT condition. We achieve the alternative on/off pump field by injecting two pump beams with a small angle to establish a standing wave. The pump field intensity has its maximum and minimum (basically zero) at the interference peak and valley, respectively. When the coupling beam is sent in as an extended uniform (Gaussian) profile and two signal beams are injected in the same directions as the pump beams, the detected signal profile exhibits interference pattern with peak experiencing gain due to the four-level Raman process and valley experiencing loss due to three-level EIT (without the pump field). This scheme is quite different from the one adopted in Ref. [35], and consequently requires a distinct laser configuration. In the current work, the two resonant signal beams (with different intensities) propagate along the $z$ direction with a small angle to establish a standing wave along the transverse direction $x$. The refractive index along the $x$ direction can be spatially engineered with the assistance of an EIT window (in the three-level configuration) generated by the co-propagating standing-wave signal field and Gaussian coupling field. Based on this periodic EIT configuration, with the standing-wave pump field turned on and tuned to completely overlap with the periodic signal field, gain and loss can be effectively generated at the bright and dark fringes of the signal field, respectively. Such periodic gain-loss pairs can be exploited to construct a spatially distributed PT-symmetric refractive index by appropriately tuning the gain/loss ratio, which can be easily manipulated by adjusting the pertinent parameters including the atomic density, the periodicity of the standing waves, and the frequency detunings and Rabi frequencies of the corresponding fields. One of the key advantages of the current system is that the gain and loss channels can be adjusted more independently, which will be easier to control the gain, loss and coupling coefficient to realize the PT-symmetric condition and study related effects. To demonstrate the characteristic features of the system, we measured the relative phase difference between two adjacent gain/loss channels by utilizing a reference interference (along the *y* direction) of the signal beam. All the experimental observations can be explained by the numerical simulations.

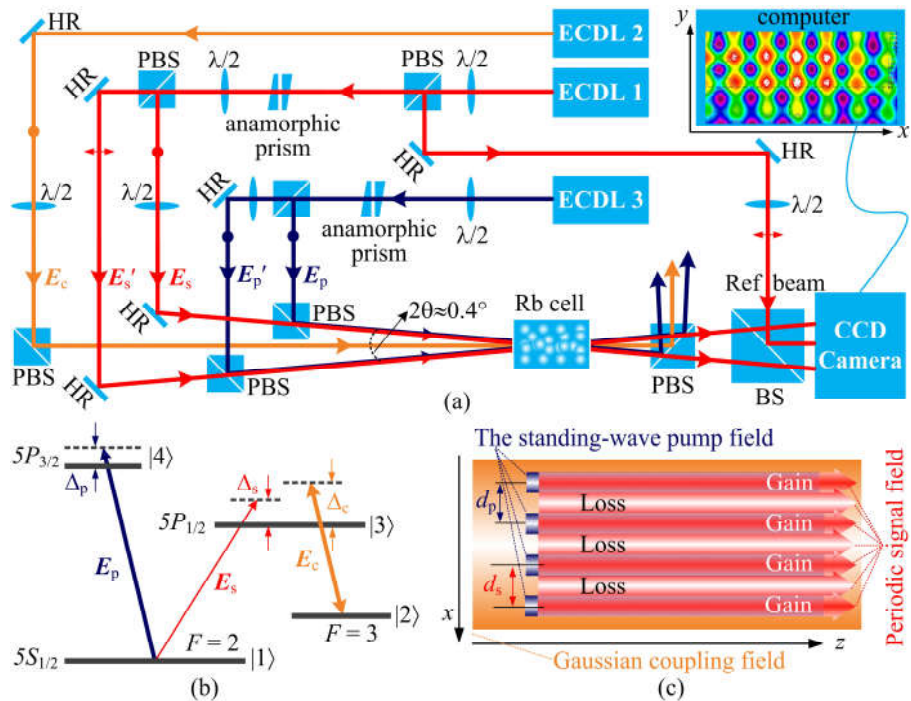

Fig.1 (a) Experimental setup. The gain-loss modulated signal beam and the phase information are detected by a CCD camera. $E_s$'&$E_s$ are signal beams from the same laser ECDL1 and the pump beams $E_p$'&$E_p$ are from ECDL3. The four laser beams ($E_s$&$E_s$' and $E_p$&$E_p$') are shaped as ellipses with approximate axial diameters of 1.5 mm and 4.5 mm, respectively, by two pairs of anamorphic prisms. In this schematic, the two signal beams separate out of the cell. In practical experiment, the two beams can keep the interference pattern for a long distance even out of the cell. The reference beam intersects with the standing-wave signal field to generate the reference interference along the $y$ direction. The double-headed arrow represents the horizontal polarization of lasers while the dot represents the vertical one. ECDL: external cavity diode laser, $\lambda/2$: half-wave plate, HR: high-reflectivity mirror, PBS: polarization beam splitter, CCD: charge coupled device. (b) The four-level $N$-type energy-level configuration for introducing gain. The three-level $\Lambda$-type subsystem $|1\rangle \rightarrow |2\rangle \rightarrow |3\rangle$ is responsible for loss. (c) The spatial arrangement of the laser fields inside the medium. The symbol $z$ represents the propagation direction of light.

The experimental scheme is shown in Fig. 1(a). The standing-wave signal field and the other two laser fields (standing-wave pump field and large-size Gaussian coupling field) propagate along the same $z$ direction to establish this periodical alternating gain and loss atomic configurations. The gain configuration is an $N$-type four-level $^{85}$Rb atomic energy-level structure (see Fig. 1(b)), consisting of two hyperfine states $F=2$ (level $|1\rangle$) and $F=3$ ($|2\rangle$) of the ground state $5S_{1/2}$ and two excited states $5P_{1/2}$ ($|3\rangle$) and $5P_{3/2}$ ($|4\rangle$), while the $\Lambda$-type subsystem $|1\rangle \leftrightarrow |2\rangle \leftrightarrow |3\rangle$ is responsible for the modifiable loss. To be specific, two elliptically-shaped signal beams $\boldsymbol{E}_s$ and $\boldsymbol{E}_s'$ [wavelength $\lambda_c=794.97$ nm, frequency $\omega_s$, Rabi frequencies $\Omega_s$ and $\Omega_s'$ ($\Omega_s \neq \Omega_s'$), respectively] from the same external cavity diode laser (ECDL1) are symmetrically arranged with respect to the $z$ axis and intersect at the center of the rubidium cell with an angle of $2\theta \approx 0.4°$ to establish a periodical signal field in the $x$ direction. Due to intensity difference between the two signal beams, the intensity of the interfering valley is not zero. The 7.5cm long cell is wrapped with $\mu$-metal sheets to shield the magnetic field and heated by the heat tape to provide an atomic density of $\sim 2.0*10^{12}$cm$^{-3}$ at 75°C. The small-angle arrangement can make the signal beam be a standing-wave field for a relatively long distance (over 20cm) along the $z$ direction.

With the frequency of the strong Gaussian coupling beam $E_c$ ($\lambda_c$=794.97 nm, $\omega_c$, $\Omega_c$) tuned to be near resonant with the transition $|2\rangle \leftrightarrow |3\rangle$, a $\Lambda$-type EIT configuration is achieved in the $|1\rangle \leftrightarrow |2\rangle \leftrightarrow |3\rangle$ subsystem. As the coupling bean is large enough to cover the standing-wave signal field, the spatially periodical EIT (along the $x$ direction) can be observed at the output surface of the cell. Namely, both the bright and dark fringes on the signal field experience adjustable loss. Furthermore, the two pump beams $E_p$ and $E_p'$ ($\lambda_p$=780.24 nm, $\omega_p$, $\Omega_p$ and $\Omega_p'$, respectively) are incident into the cell with almost the same angle of $2\theta$ to build the periodical pump field. The standing-wave pump field possesses a standard intensity distribution of the interfering wavefront, i.e., the intensity of the valley point is nearly zero. The presence of the pump field can introduce an alternating Raman gain on the signal field under the four-level N-type configuration [42]. The schematic diagram for the spatial arrangement of the three fields inside the cell is shown as Fig. 1(c), where the signal array completely overlaps with the pump array. The gain is generated when the three fields are all present (corresponding to the four-level configuration in the bright fringes) while the signal experiences loss with only the signal and coupling fields (corresponding to the three-level EIT configuration in the dark fringes).Therefore, the periodic alternating gain and loss regions for the signal beam correspond to the periodic appeared four- and three-level atomic systems. Since the signal field has orthogonal polarization from the pump and coupling fields, we can reject the pump and coupling fields with a polarization beam splitter and monitor the signal field on a charge coupled device (CCD) camera. With the gain-loss modulated periodic signal field obtained, we introduce a reference beam (from the same ECDL1 as $\boldsymbol{E}_s$) to interfere with the signal beam and demonstrate the relative phase difference (representing the internal phase difference of the eigenvectors in the system) between the gain and loss channels. Here the reference beam is coupled into the light path via a 50/50 beam splitter along the $y$ direction (see Fig. 1(a)). Both the output signal beam and the phase information are monitored by the CCD camera.

Theoretically, with the intensity of the signal field spatially modulated, the real (dispersion) and imaginary (absorption/loss) parts of the refractive index for the signal field are also modulated as a function of the transverse coordinate $x$ under the EIT condition. In the presence of the intensity-modulated (periodic) pump field, the imaginary part of the index as seen by the signal field can be alternately above (absorption/loss) and below (gain) zero along $x$.

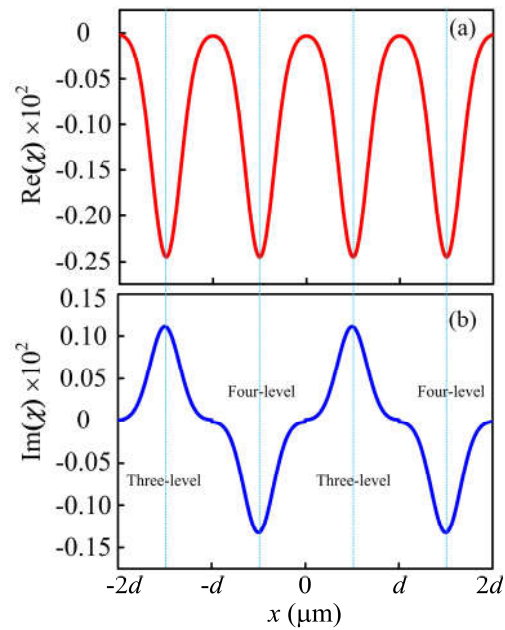

Fig. 2 Real (a) and imaginary (b) parts of the refractive index for the signal field as a function of position $x$ when the intensities of the signal field and the pump field are both spatially modified.

$\Omega_c=2\pi\times10$ MHz, $\Omega_s=2\pi\times0.2[1.5+\cos(\pi x/2)]$ MHz, $\Omega_p/2\pi=6\pi\times0.2[1+\cos(\pi x/2)]$ MHz, $\Delta_p=40$ MHz, $\Delta_c=-100$ MHz, and $\Delta_s\approx-2\pi\times15$ MHz.

In detail, based on the density-matrix equations for the four-level atomic configuration [33], the susceptibility for describing the optical response of the signal beam can be obtained through the expression $\chi=(2N\mu_{13}/\varepsilon_0 E_s)\times\rho_{31}$. Here $\rho_{31}$ is the density-matrix element for the transition $|1\rangle\leftrightarrow|3\rangle$; $N$ is the atomic density; $\mu_{ij}$ ($i, j$=1, 2, 3, 4) is the dipole moment between $|i\rangle\leftrightarrow|j\rangle$; $\varepsilon_0$ is the permittivity of vacuum and $E_f$ ($f=s, c, p$) is the electric-field amplitude of $\boldsymbol{E}_f$. Given that $n=\sqrt{1+\chi}\approx1+\chi/2$, $\chi=\chi'+i\chi''$, and $n=n_0+n_R+in_I$ ($n_0=1$ is the background index of the atomic medium), the real and imaginary parts of the refractive index can be written as

$$n_R\approx\chi'/2=N\mu_{13}/(\varepsilon_0 E_s)\times\text{Re}(\rho_{31}) \qquad \text{and} \qquad n_I\approx\chi''/2=N\mu_{13}/(\varepsilon_0 E_s)\times\text{Im}(\rho_{31})\quad.$$

Although the adopted density-matrix equations express the four-level structure, they can also be applied to describe the Λ-type three-level configuration when $\Omega_p$ is set as 0, which corresponds to the minimum intensity of the periodic pump field. To achieve the PT-symmetric conditions in the current setting, the imaginary part Im($\rho_{31}$) should be negative (gain) and positive (loss) in the respective four- and three-level configurations, while the real part Re($\rho_{31}$) should be same for the two types of atomic configurations. By making use of the advantage of this multi-parameter tunable system, in which the loss and gain can be independently adjusted by the parameters (mainly the Rabi frequency $\Omega_f$ and frequency detuning $\Delta_f$) of the coupling and pump beams, we can construct the desired refractive index satisfying $n(x)=n^*(-x)$ as shown in Fig. 2. Here the frequency detunings for the signal, coupling, and pump fields are defined as $\Delta_s=\omega_{31}-\omega_s$, $\Delta_c=\omega_{32}-\omega_c$, and $\Delta_p=\omega_{41}-\omega_p$, respectively.

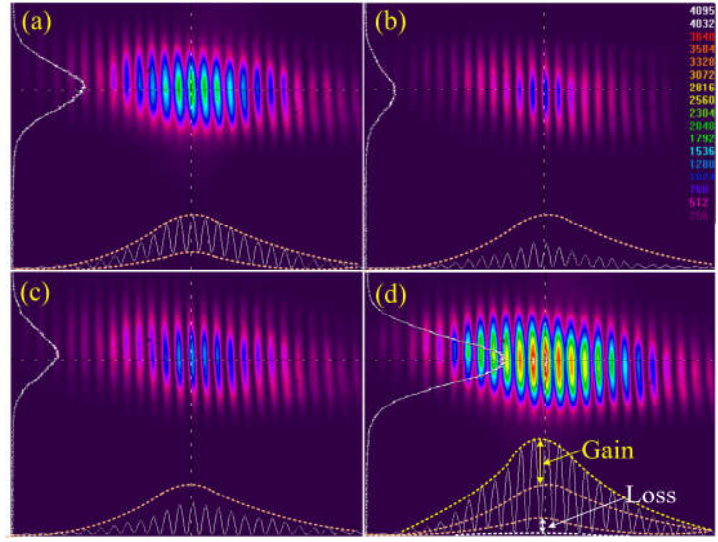

Fig. 3 Detected signal field under various conditions. (a) Observed image and intensity profiles of the signal field $\boldsymbol{E}_s$ without interacting with atoms. (b) Signal field after propagating through the atomic medium under the resonant condition (simple atomic absorption). (c) Reduced absorption with generated EIT when the Gaussian coupling field $\boldsymbol{E}_c$ is turned on. (d) Simultaneous alternative gain and loss profiles on $\boldsymbol{E}_s$ with both $\boldsymbol{E}_c$ and pump field $\boldsymbol{E}_p$ turned on.

Figure 3 depicts the detected signal field on the CCD camera under different conditions. First, with the frequency detuning of the signal field $\boldsymbol{E}_s$ far away from the atomic resonance, one sees the simple interference fringes in Fig. 3(a). As the signal frequency tuned to be near resonant with the transition $|1\rangle\leftrightarrow|3\rangle$, it can be dramatically absorbed (Fig. 3(b)). When the large Gaussian coupling field $\boldsymbol{E}_c$ is turned on, it induces EIT on the signal field and reduces its absorption (Fig. 3(c)) under the two-photon resonance satisfying $\Delta_s-\Delta_c=0$. The amount of residual absorption is determined by the coupling beam parameters. Then, by injecting the pump field into the cell, simultaneous alternative gain and loss channels with a controllable contrast are obtained on the signal field by carefully arranging the coupling- and pumping-field parameters. Inside the cell, the two standing-wave fields with the same periodicity completely overlap with each other. The presence of the pump field can couple a four-level N-type system and therefore can lead to an amplification on the signal field without population inversion ($\rho_{11} > \rho_{33}$) [42]. Notice that the intensity of the pump field at the dark-fringe channels is almost zero, which indicates that a three-level EIT configuration (with only $E_c$ and $E_s$) exists in dark-fringe channels. As shown in Fig. 3(d), one can clearly find the alternative gain (bright fringes with three fields on) and loss (dark fringes with EIT) in the middle region of the periodical field by comparing it with the original intensity profile of the signal field shown in Fig. 3(a).

By taking advantages of the atomic coherence, the ratio between the induced gain and loss can be easily controlled by tuning experimental parameters. The ratio $\gamma_G/\gamma_L$ ($\gamma_G$ and $\gamma_L$ representing the gain and loss coefficients, respectively) for describing the properties in the non-Hermitian system can be directly measured from the generated gain and loss components. In the current atomic configuration, the intensity for the detected signal field can be written as $I = I_0 - I_0 e^{-aL} \approx I_0 - I_0(1-aL)$, where $a = (2\pi/\lambda_s)\chi''$, where $I_0$ is defined as the initial intensity of the signal field and $L$ as the cell length. The loss or gain is determined by the sign of $\chi''$, namely, the negative (positive) $n_I$ represents the gain (loss). Consequently, we have $I \propto \chi'' = 2n_I$, which means that the ratio between gain and loss intensities can be equivalent to the ratio $\gamma_G/\gamma_L$. Figure 4(a) gives the original intensity profile of the standing-wave signal field, and Figs. 4(b1)-(b5) shows the typical evolution of the ratio $\gamma_G/\gamma_L$ versus the detuning $\Delta_c$ of the coupling field with the signal-field detuning tuned to be near resonant with the transition 5S$_{1/2}$, $F$=3→5P$_{1/2}$, $F'$=3 of $^{85}$Rb. With the detuning $\Delta_c$ tuned from the far detuned points ($\Delta_c$=±20 MHz) to the resonant point ($\Delta_c$=0), the gain can be greatly enhanced, which can be attributed to the better satisfaction of the two-photon resonant condition $\Delta_s$−$\Delta_c$=0.

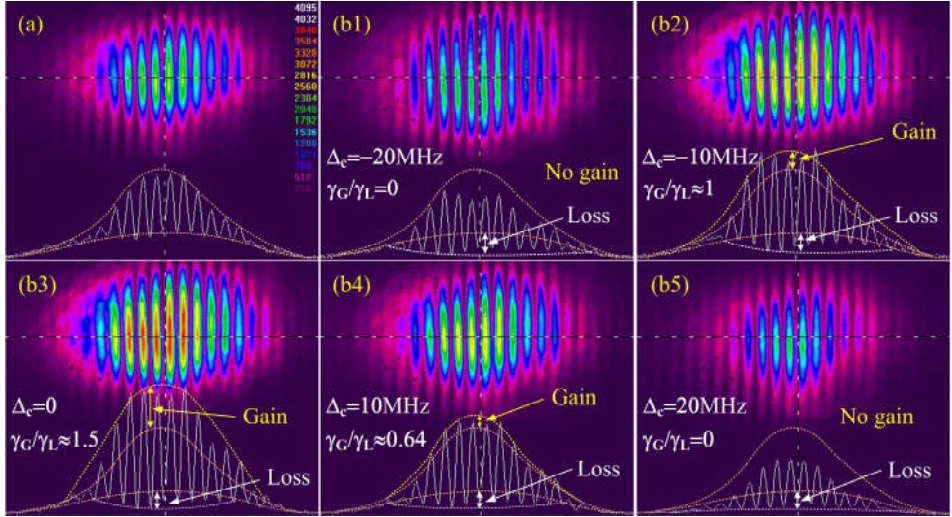

Fig. 4(a) The initial intensity profile of $\boldsymbol{E}_s$ without absorption. (b1)-(b5) Typical evolution of the gain/loss ($\gamma_G / \gamma_L$) ratio versus the coupling detuning $\Delta_c$.

The results in Fig. 4 show that the ratio $\gamma_G / \gamma_L$ can be in the range of 0 to 1.5, covering the requirement of $\gamma_G / \gamma_L$=1 (balanced gain and loss) for exact PT symmetry. For an unbalanced gain-loss profile (namely, $\gamma_G \neq \gamma_L$), it can be mathematically transformed into an exact PT-symmetric one ($\gamma_G = \gamma_L$) by using of the gauge transformation [6, 7] in Wick space, and be understood as "quasi-PT symmetry" [43]. The characteristic eigenvalue pattern of a quasi-PT symmetric Hamiltonian becomes with respect to a new offset center instead of the original zero point [44]. For the cases below the threshold of symmetry breaking, the PT-symmetric system and quasi-PT-symmetric one can experience the same dynamics behaviors. For the case above the threshold, the offset of the symmetric point in a quasi-PT-symmetric system makes the dynamics different from its PT-symmetric counterpart. Fortunately, the introduced difference can be eliminated by use of $z$-dependent power normalization. Consequently, an unbalanced gain-loss case can also demonstrate the dynamics behaviors similarly to the exactly balanced case.

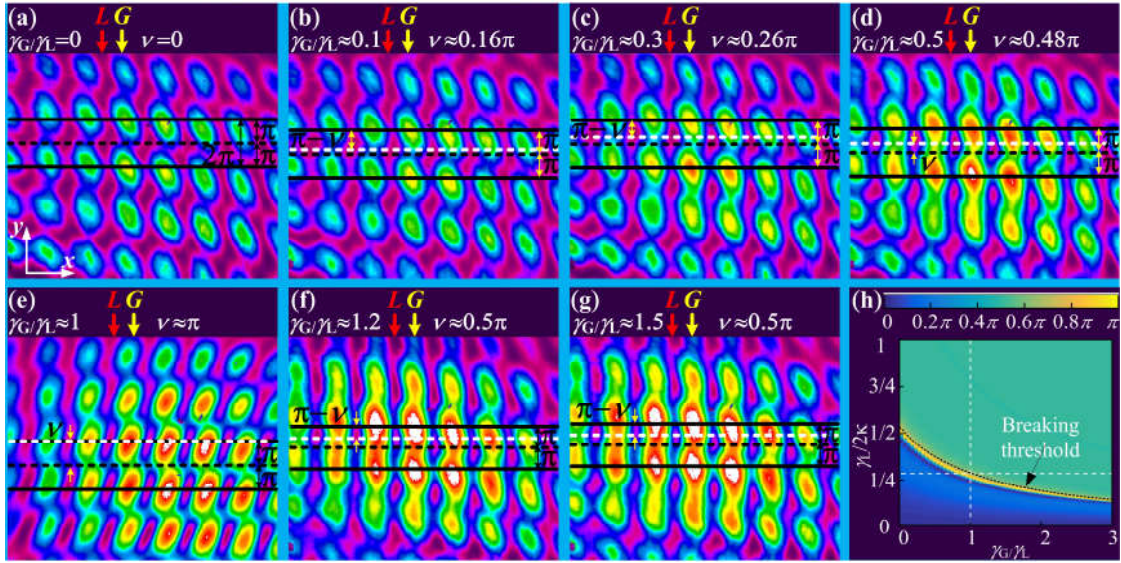

Fig. 5 Measured phase difference between two neighboring gain and loss channels, which are marked by $G$ and $L$, respectively. (a) The phase difference $v=0$ for the case without gain. The double-sided arrow between the two black solid lines marks the distance between two adjacent fringes along the $y$ direction and its length is defined as $2\pi$. (b)-(e) The measured $v$ at different gain/loss ratios for the cases below the PT-symmetry breaking threshold. The white dashed line marks the center of dark square in the loss channel, and the distance between the white and black dashed lines represents the relative phase difference $v$ between the gain and loss channels. (f)-(g) The measured $v$ is fixed at $\pi/2$ when PT symmetry is broken. (h) The simulated phase difference with 10 gain-loss channels coupled. Term $\kappa$ is the coupling coefficient between two adjacent gain and loss channels.

The dynamics behaviors of the established non-Hermitian system below and above the threshold are demonstrated by the relative phase difference ν between two neighbouring gain and loss channels. A reference beam from the same laser as the probe beam is introduced to interfer with the generated gain-loss lattice, and the reference pattern looks like a two-dimentional lattice with bright (in gain channels) and dark (in loss channels) squares. Figure 5(a) shows the case without gain, namely, the phase difference is ν=0. In Fig. 5(a), we add two black solid lines to mark the center of two adjacent fringes of the reference interference along the $y$ direction, and the distance between them is defined as $2\pi$. The distance mentioned in the current work is along the $y$ direction. The black dashed line locates at the middle between the two solid lines and just marks the center of the dark square in the case of no gain. Consequently, the distance between the marked bright square (marked by the upper black solid line) and dark square is $\pi$. With the gain introduced, the relative distance between the dark and bright regions can be modified by changing the gain/loss ratio.

As shown in Figs. 5(b)-5(e), the relative phase difference can increase from $0.16\pi$ to $\pi$ with the gain/loss ratio adjusted from 0.1 to 1. The phase difference is obtained by measuring the relative distance between the black and white dashed lines. The added white dashed line represents the center of the shifted dark square. By proportionally comparing the measured distance with the length of the shorter double-sided arrow representing $\pi$, one can obtain the value of the phase difference. The phase difference directly reflects the internal phase difference of the eigenvectors in a PT-symmetric system, while the eigenvalues are the basic criteria to determine whether the system is operating below or above the symmetry breaking threshold. Theoretically, the corresponding phase difference below and above the threshold should be different [6]. For the case below the phase-transition point, the relative phase differences ν can increase (from the initial values 0) with the gain/loss ratio. For the cases at and above the phase-transition point, the phase differences can constantly be π/2 even increasing the gain. The phase information under the condition of PT-symmetry breaking is given at points $\gamma_G/\gamma_L$=1.2 and 1.5, as shown in Figs. 5(f) and 5(g), respectively. Actually, when the ratio $\gamma_G/\gamma_L$ is increased to 1.2, the phase difference jumps from π ($\gamma_G/\gamma_L$=1) to ν=π/2. With the ratio further increased to $\gamma_G/\gamma_L$=1.5, the measured phase difference can still be π/2.

Figure 5(h) theoritically predicts the evoluiton of the phase difference in a PT-symmetric lattice with 10 gain-loss channels coupled. The dynamic behaviors of the system is calculated based on the paraxial wave equation $i\,\partial E/\partial z + \partial^2 E/\partial x^2 + V(x)E = 0$, where $V(x)$ is the potential of the periodic gain-loss structure and $E$ is the electric field envelope in each channel [3, 6, 30, 35]. The theoritical simulations can adovcate our experimantal reults. Particularly, for the points below but very close to the threshold, the phase difference can experience the π value in a very small range, which is also certified by the experiment.

In summary, a PT-symmetric optical lattice with controllable gain/loss ratio and the phase transition at the symmetry breaking threshold are experimentally demonstrated based on spatially alternating coherently-prepared four-and three-level atomic configurations in the same [85]Rb atomic sample. The gain-loss profiles are generated by exploiting the modified absorption and active Raman gain only with the assistance of EIT technique. Compared to the previous work based on diffraction [35, 45], the current scheme based on easily acquired EIT is much easier to be operated in experiment. Also, the parameters in the two different atomic configurations responsible for gain and loss can be modified separately under certain conditions. By slightly adjusting the angle for establishing the standing-wave fields, the coupling efficiency $\kappa$ between the gain and loss channels can be promisingly modified but without changing the gain-loss ratio if every beam is a plane wave instead of the current Gaussian profile. Consequently, by taking the advantages of the large available parametric spaces in atomic configurations, the newly demonstrated scheme can provide a more flexible platform to investigate the diverse effects resulting from the interplay between non-Hermitian Hamiltonian and nonlinear effects [46, 47], and a variety of properties resulting from the intriguing beam dynamics in non-Hermitian optical lattices.

## Acknowledgements

This work was supported in part by National Key R&D Program of China (2017YFA0303700), National Natural Science Foundation of China (61605154, 11474228), Key Scientific and Technological Innovation Team of Shaanxi Province (2014KCT-10), Natural Science Foundation of Shaanxi Province (2017JQ6039, 2017JZ019) and China Postdoctoral Science Foundation (2016M600776, 2016M600777, 2017T100734). L. Y. is sponsored by Fundamental Research Funds for the Central Universities (HEUCFJ170402).